# High Dynamic Range Imaging Using a Deformable Mirror for Space Coronography

F. Malbet[1], J.W. Yu, and M. Shao

Jet Propulsion Laboratory,
California Institute of Technology,
Mail Stop 306-388,
4800 Oak Grove Drive,
Pasadena, CA 91109-8099, USA
Electronic mail: malbet@huey.jpl.nasa.gov

## ABSTRACT

The need for high dynamic range imaging is crucial in many astronomical fields, such as extra-solar planet direct detection, extra-galactic science and circumstellar imaging. Using a high quality coronograph, dynamic ranges of up to $10^5$ have been achieved. However the ultimate limitations of coronographs do not come from their optical performances, but from scattering due to imperfections in the optical surfaces of the collecting system. We propose to use a deformable mirror to correct these imperfections and decrease the scattering level in local regions called "dark holes". Using this technique will enable imaging of fields with dynamic ranges exceeding $10^8$. We show that the dark-hole algorithm results in a lower scattering level than simply minimizing the RMS figure error (maximum-strehl-ratio algorithm). The achievable scattering level inside the dark-hole region will depend on the number of mirror actuators, the surface quality of the telescope, the single-actuator influence function and the observing wavelength. We have simulated cases with a $37 \times 37$ deformable mirror using data from the *Hubble Space Telescope* optics without spherical aberrations and have demonstrated dark holes with rectangular and annular shapes. We also present a preliminary concept of a monolithic, fully integrated, high density deformable mirror which can be used for this type of space application.

*Subject headings:* adaptive optics – active optics – coronography – scattering – extra-solar planets – circumstellar imaging

---

[1]This work was performed while the author held a National Research Council-JPL Research Associateship Current address of the author: Laboratoire d'Astrophysique, BP 53X, F-38041 Grenoble cedex, FRANCE (e-mail: malbet@gag.observ-gr.fr)

## 1. INTRODUCTION

High dynamic range observations are necessary for many astronomical applications that require the detection of very faint objects close to a bright object. Coronography can reduce the diffracted light coming from the bright on-axis object to a very high level. But small figure errors on the collecting mirrors will produce scattered light. In a very good coronographic system, the dynamic range will not be limited by the light level from the diffraction pattern of the bright object, but rather by the scattering from figure errors in the mirror surface.

Recently, adaptive optics have helped coronographs reach very high rejection rates (Malbet 1992, Golimowski 1993) by stabilizing the image and removing the effect of the atmospheric jitter. In space, where atmospheric correction is not necessary, coronography can be even more efficient. However, the quality of the mirror becomes very important and scattered light is the main limitation to high dynamic range observations. Since the scattering close to the optical axis comes from the figure errors of the mirror with large spatial scale, adaptive optics for space could provide a method of decreasing the on-axis scattering level by correcting mirror imperfections with a deformable mirror. This paper describes how to decrease the scattering level in some local regions called "dark holes" using a new kind of deformable mirror.

Deformable mirrors for astronomy are designed to compensate for the wavefront error caused by atmospheric turbulence. Usually the facesheet is thin compared to the diameter of the mirror, and, the actuators are located between the backup plate and the mirror facesheet in order to deform the reflecting surface. In astronomy, distinctions are made between active optics systems and adaptive optics systems. In active optics systems, the deformable mirror is the collecting aperture. The wavefront correction is not fast (typically 0.1-0.001 Hz) and corrects only the first-order aberrations (coma, astigmatism,...) of the telescope. Adaptive optics systems use deformable mirrors which are much smaller (10-20 cm) at an image of the entrance pupil. There are many more actuators (20-250) which typically are piezo-electric stacks moved by modulating an electric field generated from a high-voltage supply ($\approx 500$ V). In this case, the closed-loop bandwidth is much higher, up to 1 kHz.

The application presented in this paper is mid-way between active optics and adaptive optics. As in the case of adaptive optics system, a large number of active elements (actuators) are needed, however since the figure errors are static we only need to correct at a low bandwidth.

The future instruments which could benefit from this type of deformable mirror are mainly space-based telescopes. With such a device, the *Hubble Space Telescope* (*HST*) would be able to image a planet around a solar-type star located at 10 pc (Malbet, Shao and Yu 1994). Likewise the performances of the proposed *Astrometric Imaging Telescope* (*AIT*, Pravdo et al. 1994) would be increased.

Section 2 of the paper presents the astrophysical objectives of a space-based adaptive optics coronograph. Section 3 describes in more details the principle of this technique and its fundamental limits. Section 4 show the details of the algorithm used to reduce the scattering in the dark-hole regions and Sect. 5 presents simulation results. The limitations of the algorithm are discussed in Sect. 6. Finally in Sect. 7, a conceptual design and a method of fabricating the space deformable mirror is discussed.

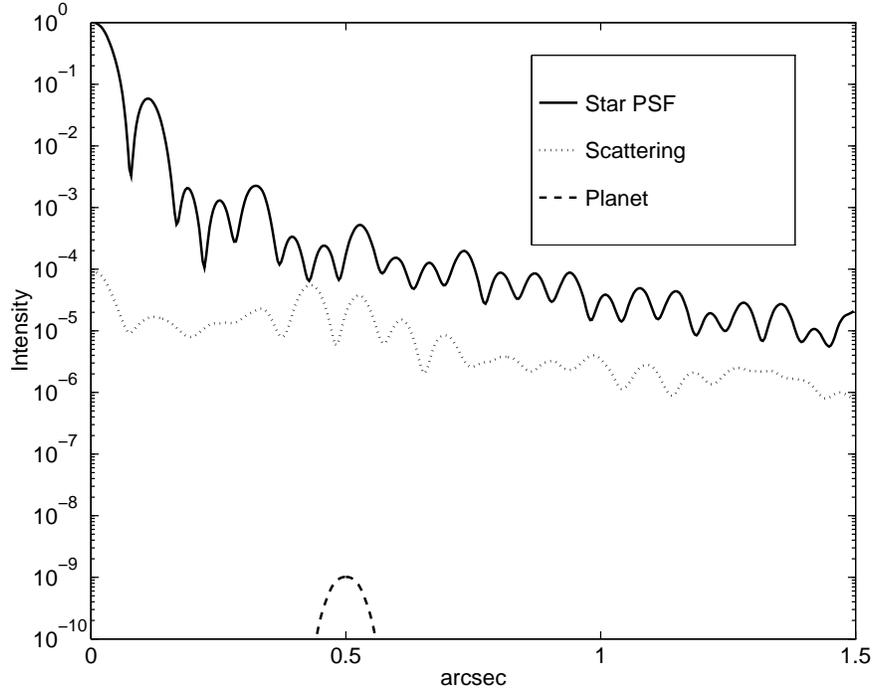

Fig. 1.— Diffraction pattern of star and planet as observed by HST telescope at 0.8 $\mu$m. The dashed line represents the scattering level due to small figure errors on the *HST* mirror ($\lambda/60$ RMS phase error). The radial profiles have been azimuthally averaged.

## 2. SCIENCE OBJECTIVES

### 2.1. Extra-Solar Planet Detection

The key problem in the detection of planets around other stars is that the planet is much fainter than the star. In the case of a Sun-Jupiter system the flux ratio is about $10^{-9}$. Figure 1 shows the diffraction patterns of the Sun and Jupiter located 10 pc away and the scattered light level from *HST*. The *HST* data was taken from a report by Roddier and Roddier (1990) and is displayed on Fig. 2. Phase-retrieval techniques were used on 8-9 WF-PC-1 images during the *HARP* (*Hubble Aberration Retrieval Program*) program to assess the spherical aberration on *HST*.

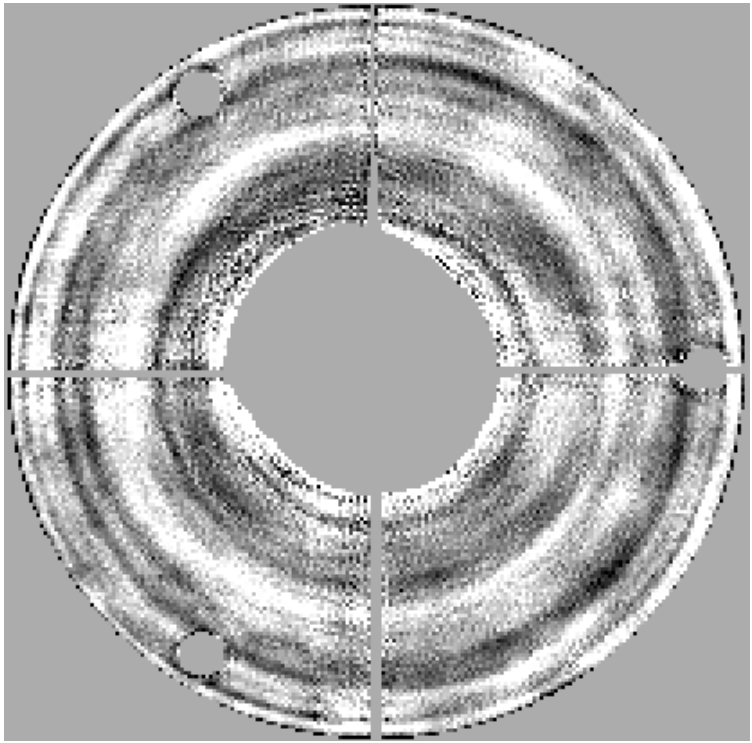

Fig. 2.— *HST* phase map from Roddier and Roddier (1990) without the spherical aberration. The computed RMS phase error is 1.3 $\mu$m in amplitude.

The scattering level is computed at $\lambda = 0.8$ $\mu$m by taking the difference between the diffraction pattern generated from *HST* phase map and that generated from a perfect phase map. This is similar to using a rotating shearing interferometer in order to null out the light from the star. In our simulations, the phase error due to spherical aberrations has been removed from the *HST* phase map. The result can be compared to the model of Brown & Burrow (1990). Taking the Strehl ratio for HST low-spatial-frequency aberrations equal to about 0.95 for $\lambda = 0.8$ $\mu$m (it is almost constant between 0.5 $\mu$m and 1.0 $\mu$m), the asymptotic radial profile of the normalized intensity at $\lambda = 0.8$ $\mu$m is proportional to $6 \times 10^{-5} \theta('')^{-3}$ for the PSF and $4 \times 10^{-6} \theta('')^{-2.19}$ for the scattering level. At angular distance $\theta = 1''$, the match is almost perfect. The match is not as good at $\theta \leq 0.5''$ certainly because of the presence of the three pads and the spider.

For a planetary system located 10 pc away, Jupiter would be separated by 0".5 from the sun. In the case of *HST*, the planet signal would be $10^5$ dimmer than the intensity due to the diffraction wing of the star. Assuming a total transmission of 35% (80% for the primary and secondary aluminum mirror, 90% for the detector optics and 0.6 of quantum efficiency), a 3.3 m$^2$ collecting area, at $\lambda = 0.8$ $\mu$m with a 0.4 $\mu$m bandwidth, *HST* would receive $8 \times 10^8$ photons s$^{-1}$ for a sun-like star and 0.8 photons s$^{-1}$ for the Jupiter-like planet. Therefore the signal-to-noise ratio for 1 s of integration time would be $3 \times 10^{-3}$. To achieve an SNR of 1 would require an extremely long integration time of 35 hours.

The main obstacle to observing extra-solar planets is the photon noise from the star diffraction wing. Using a coronograph can reduce this contribution by a factor $10^2$ to $10^6$. However in the case of *HST*, the scattered light will limit the star background to still be $10^4$ times brighter than the planet. This will lead to an integration time of 3h 30m for an SNR of 1 and 30 h for an SNR of 3.

In order to get a reasonable SNR (SNR = 5) in a reasonable amount of integration time ($t = 1$ h) the background reduction should be no more than 120 times brighter than the planet or at least $1.2 \times 10^{-7}$ dimmer than the intensity of the central peak. Due to losses in the additional coronograph optics, we estimate that the scattering level needs to be decreased to $3 \times 10^{-8}$ of the the star peak intensity (factor 4).

In the following sections, we will show that this level of noise reduction is achievable in small regions of the focal plane. For extra-solar planet detection, these zones will be chosen by focusing in on the expected position of the planet. As an example in our solar system, we expect to find giant planets between 5 and 20 AU, which correspond to $0\rlap{.}''5$ to $2''$ on the sky for a star located at 10 pc.

## 2.2. Circumstellar Imaging

Extra-solar planet detection is part of the process of imaging the circumstellar surroundings of stars. Nevertheless in most of the star systems that we presently know of, the circumstellar material detected is not yet planetary. This material fall into three different categories: the dust and/or gas envelopes, the accretion disk or envelopes and the ejected material.

Young stellar systems, protoplanetary disks, stellar jets, reflection nebulae, gas and dust reservoirs can be directly or indirectly detected. In the case of the star HL Tauri, the observed flux from the different jets are $I = 22$ (Mundt et al. 1990) compared to a total flux of $I = 12.5$ (Herbig and Bell 1988), which comes mainly from the star. This results in a flux ratio of $1.6 \times 10^{-4}$. Moreover, compared to other young stars, HL Tau is an object where the jets and related features are relatively bright. Direct imaging of similar objects can only be accomplished only if very high dynamic range sensing is possible.

For $\beta$ Pictoris, the flux distribution of the disk is between 13 mag arcsec$^{-2}$ at radius of 40 AU and 21 mag arcsec$^{-2}$ at radius of 480 AU (Golimowski, Durrance, and Clampin 1993) in the $R$-band compared to a total flux of 3.72 mag in $R$-band. The disk features are therefore between $2 \times 10^{-4}$ and $1.5 \times 10^{-7}$ times fainter than the star (assuming the optimistic case that the star intensity is spread uniformly over a 1 arcsec$^2$ area) for disk radii between $2\rlap{.}''5$ to $25''$ ($D = 16.4$ pc). For a similar star located 10 times further, an optical system with a dynamic range of about $10^8$ and subarcsecond angular resolution would be required.

Other targets for a space-born active optics coronograph are the circumstellar surroundings of supergiant. Origin of the stellar wind, as well as the morphology of the interaction between the stellar wind and the surrounding medium could also be studied if very high dynamic range is achievable simultaneously with high spatial resolution.

### 2.3. Extragalactic Science

Observations of active galactic nuclei (AGN) require very high dynamic range imaging near the nuclei (< 1 kpc). With high dynamic range imaging technique, the radiation from the optical jets (Miley, 1981), scattered emission from ionized gas and stars of different ages can be studied.

Other interesting observations include imaging galaxy fields around quasars and quasar fields around foreground galaxies in order to study their global environment. Yee et al. (1986) have shown that the galaxy count around quasars deviates significantly from Poisson statistics. In a sample of quasars with Gunn magnitude $15 \leq r \leq 17$, the galaxy count deviation occurs for galaxies with magnitude greater than 21. With a dynamic range of $10^{-6}$, we could measure objects up to $r \leq 27$ within $1''$ from the quasar. Wu (1994) discussed the interest of quasar and galaxy counting in small sky fields. As shown in Wu (1994), imaging the surroundings of galaxies within a $1''$ radius may reveal a quasar enhancement factor larger than 5 times the normal value. Such observations would also be useful to detect quasar companions that might trigger the quasar activity (Yee 1987).

## 3. PRINCIPLE

### 3.1. Origin of Scattered Light

Scattering originates from small figure errors on the collecting optics. The small figure errors change the impulse response (or point spread function, PSF) of the telescope as compared to a perfect system. In order to estimate the scattering level, we need to subtract the perfect PSF from the actual PSF. The subtraction can be done either on the intensity or the amplitude. Amplitude subtraction can be achieved by using a pupil-plane shearing interferometer with a perfect mirror in one arm and the imperfect optics in the other arm. The image formed is due to the figure errors of imperfect optics and shows the amount of scattered light as function of field position.

Consider an optical system with a small phase error $\delta\phi(u,v)$. The resulting intensity from the shearing interferometer would be:

$$I(x,y) = |\text{PSF}(x,y) * \text{FT}\left(\exp(i\delta\phi(u,v)) - 1\right)|^2, \tag{1}$$

where PSF is the ideal point spread function, i.e. with no phase errors, $*$ is the convolution operation and FT is the Fourier Transform operator.

If the figure errors are small ($\delta\phi(u,v) \ll 1$) then the scattering level is proportional to the Fourier transform of phase error:

$$I(x,y) \approx 4 \left|\mathrm{PSF}(x,y) * \mathrm{FT}(\delta\phi)\right|^2. \tag{2}$$

Consider a simple example, where the phase error is sinusoidal:

$$\delta\phi(u,v) = \frac{2\pi h_0}{\lambda} \sin\left(\frac{2\pi u}{l_0}\right), \tag{3}$$

where $h_0$ is the amplitude of the deformation of the mirror, and $l_0$ is the spatial period of the deformation in the $u$-direction. The resulting scattering is the sum of two PSFs located symmetrically from the optical axis at an angular distance corresponding to the spatial frequency of the deformation, $\lambda/l_0$.

$$I(x,y) = \left(\frac{\pi h_0}{\lambda}\right)^2 \left[\mathrm{PSF}(x-l_0) + \mathrm{PSF}(x+l_0)\right]. \tag{4}$$

Note that this is true only if $h_0 \ll \lambda$. The intensity of the resulting scattering is about $(\pi h_0/\lambda)^2$. The result of a simulation at 0.8 $\mu$m is shown on Fig. 3 for the case of a deformation of magnitude $\lambda/50$ (i.e. $h_0 = 0.016$ $\mu$m). We took a spatial period of $l_0 = 0.35$ m corresponding to an angular distance of 0″.41.

One can then generate a sinusoidal deformation with a deformable mirror and decrease the scattering at this spatial frequency. In general, one can choose the shape of the deformable mirror to decrease the scattering at particular positions (called "dark holes") in the image: the algorithm that computes the actuator strokes is described in sections 3.3. and 4..

### 3.2. Maximum Strehl Ratio

A commonly used algorithm in adaptive and active optics is to maximize the Strehl ratio. The Strehl ratio is the ratio between the amplitude of the actual PSF and the perfect PSF. Since 1 minus the Strehl ratio is proportional to the square of the RMS phase error over the pupil, to maximize the Strehl ratio we only need to move the actuators so that their strokes minimize the RMS phase error over each actuator surfaces. This is achieved by setting each actuator with a stroke opposite to the average phase error over that actuator surface.

Figure 4 shows the result of the "max Strehl" algorithm for the case of *HST* optics and with different numbers of actuators across the diameter: $9 \times 9$ actuators (48 actually used on the mirror), $19 \times 19$ actuators (210 used), $37 \times 37$ actuators (644 used), and $51 \times 51$ actuators (1224 used). The more actuators, the lower the scattering level. The present state of the art limits the number of actuators to less than 1000 (see section 7.), which corresponds to about 40 actuators across the pupil. For our simulations, we used an actuator density of $37 \times 37$.

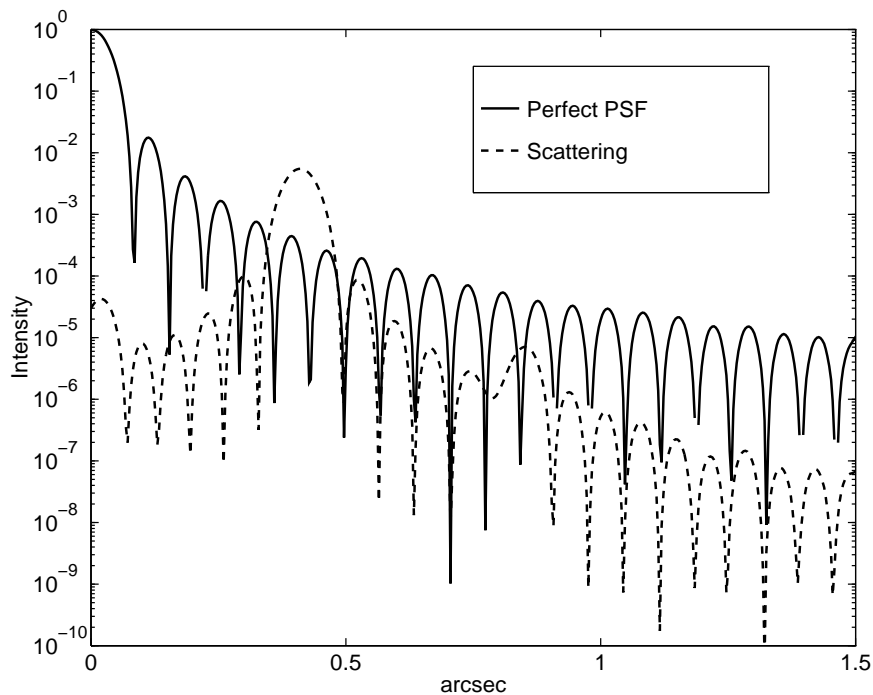

Fig. 3.— Scattering from a sinusoidal phase error on a circular pupil. The simulation has been computed at 0.8 $\mu$m for the case of a deformation with a magnitude of $\lambda/50$ (i.e. $h_0 = 0.016$ $\mu$m) and a spatial period corresponding to an angular distance of $0''\!.41$ (i.e. $l_0 = 0.35$ m). The PSF is different from the one on Fig. 1 because the reference mirror is a circular mirror with a perfect surface.

### 3.3. Scattered Light Suppression in Local Regions

The "Max Strehl" algorithm only corrects the low frequency components of the phase error and therefore only minimizes the scattering near the optical axis. Since the deformable device is to be used for very high dynamic range imaging, the on-axis light is not important because it emanates from the central bright object. It is more useful to minimize the scattered light away from the optical axis.

As shown previously, one can control the scattering in one point in the image plane using a sinusoidal deformation. In general, we can also generate a suitable deformation, which will cancel out the scattering in one or several defined regions which are called dark holes. Section 4. describes the algorithm developed to minimize the scattering in the dark holes and Section 5. shows the simulations result.

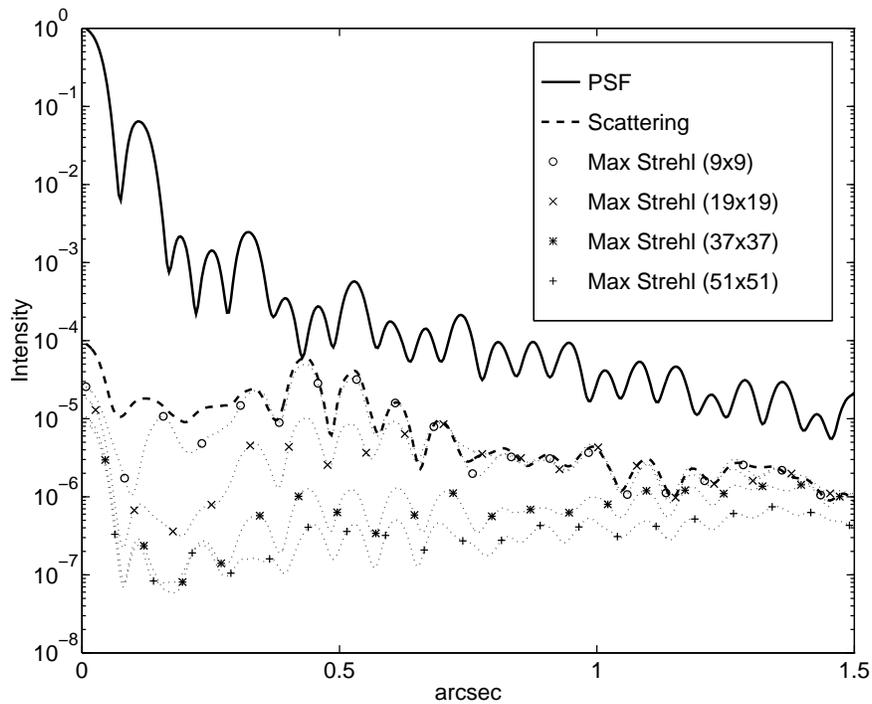

Fig. 4.— Scattering level after correction by deformable mirrors with different number of actuators. The solid line is the image PSF. The dashed line is the scattering level without any correction. The dotted lines are the scattering level reached with different numbers of actuators. From top to bottom: $9 \times 9$ actuators (48 actually used), $19 \times 19$ actuators (210 used), $37 \times 37$ actuators (644 used), and $51 \times 51$ actuators (1224 used). The radial profiles have been azimuthally averaged.

The deformable mirror cannot arbitrarily cancel scattered light in any region, and is limited by the number of active elements in the mirror. In order to correct the scattering at one point, we need to generate the sinusoid corresponding to this position at the pupil. For a point located at $\theta$ radians from the optical axis, we need to generate a sinusoid with a period $\lambda/\theta$. Due to the Nyquist limit, with $n$ actuators across the pupil and $D$ the pupil diameter, one can only generate sinusoid with period larger than $2D/n$. The pupil is finite, so that one cannot generate sinusoids with periods larger than $2D$. Since the period has to be within $2D/n$ and $2D$ in the pupil plane, the dark-hole points are limited to be within $\lambda/2D$ and $n\lambda/2D$ in radius in the focal plane. This roughly corresponds to an outer limit of $n$ Airy rings from the center.

Since Eq. (1) involves a convolution by the system PSF, the intensities between two close points in the focal plane are correlated. This can be seen by considering that the effect of a sinusoid generated by a deformable mirror does not cancel out a single point but rather an Airy disk of radius $\lambda/D$. The number of dark-hole points will be important in defining the metric in the dark-hole algorithm and the ability for the algorithm to find a solution.

When the phase errors are small, Eq. (1) can be approximated by Eq. (2) and the scattering corresponds to the Fourier transform of a purely imaginary function. Therefore the scattering from small phase errors is axisymmetric and the effect of the deformation will be symmetrical about the optical axis. In order to use only small perturbations of the deformable mirror, the dark holes must be chosen to be symmetric about the optical axis. To create asymmetric dark holes would require much larger actuator strokes typically on the order of the wavelength.

## 4. DARK-HOLE ALGORITHM

### 4.1. Dark-Hole Metric

In order to achieve the minimum scattering in the regions called "dark holes", we define a metric to be minimized. The scattered intensity in the focal plane is:

$$I(x,y) = \left| A_0(x,y) * \mathrm{FT}\left\{ \exp\left( i\sum_{k=1}^{M} a_k \sigma_k(u,v) \right) - \exp\left(-i\phi_0(u,v)\right) \right\} \right|^2, \quad (5)$$

where $A_0(x)$ is the amplitude response function of the optical system with all the actuator set to zero,

$$A_0(x,y) = \mathrm{FT}\left( \Pi(u,v) e^{i\phi_0(u,v)} \right), \quad (6)$$

and $\phi_0(u,v)$ is the actual phase error of the optical system. $\Pi(u,v)$ is the pupil transmission function of the telescope (i.e. $\Pi(u,v) = 1$ inside the telescope aperture and $\Pi(u,v) = 0$ outside). $a_k$ and $\sigma_k(u,v)$ are the stroke, i.e. the length of the actuator displacement, and influence function, i.e. the shape of the mirror induced by a stroke unity, of the $k^{th}$ actuator. $M$ is the number of actuators that are used to correct the wavefront. For the problem to be underdetermined, the number of dark-hole points, $P$, where we try to minimize the intensity should be less than $M$. The metric is then defined by summing the intensity over all the dark-hole points:

$$\chi^2 = \sum_{p=1}^{P} I(x_p, y_p) \quad (7)$$

### 4.2. Linear Solution

Since the phase errors are small, we can linearize Eq. (5). This leads to a system of $P$ linear equations:

$$\sum_{k=1}^{M} a_k f_k(x_p, y_p) = i(f(x_p, y_p) - f_0(x_p, y_p)), \text{ for } 1 < p < P, \quad (8)$$

where the functions $f_k$, $f$ and $f_0$ are defined by:

$$f_k(x,y) = \text{FT}\left\{\sigma_k(u,v)\Pi(u,v)e^{i\phi_0(u,v)}\right\} \tag{9}$$

$$f(x,y) = \text{FT}\left\{\Pi(u,v)e^{i\phi_0(u,v)}\right\} \tag{10}$$

$$f_0(x,y) = \text{FT}\left\{\Pi(u,v)\right\}. \tag{11}$$

$f(x,y)$ is the actual PSF, $f_0(x,y)$ is the perfect PSF, and $f_k(x,y)$ is the effect of moving the $k^{th}$ actuator on the scattering wavefront amplitude. $A_0(x,y) = (f(x_p,y_p) - f_0(x_p,y_p))$ is the scattering wavefront amplitude of the actual optical system in the focal plane when no actuator strokes are applied.

By calculating the inverse or the pseudo-inverse matrix, we are able to solve for $a_k$, the actuator strokes to be applied. However the linear solution has some drawbacks. The linear solution results in the best set of actuator strokes which minimizes the linear part of Eq. (7). Consequently second order effects become predominant. Our simulations show that we can only obtain good minimization of the metric over dark holes with small areas ($M \gg P$). Consequently, we decided to solve the more general non-linear equations which take into account the effect of higher order terms.

### 4.3. Non-Linear Solution

In the non-linear solution, the metric Eq. (7) is minimized using a Levenberg-Marquardt non-linear least-square program. In order to calculate the $\chi^2$ and also the gradient required for the algorithm, we need to compute $A(x_p,y_p)$ the wavefront amplitude for the points in the dark holes and also the partial derivatives $\partial A/\partial a_k(x_p,y_p)$, for a particular set of actuator strokes. That means that for each iteration we need to perform $M+1$ Fourier transforms in order to compute the image at the focal plane and determine the next guess of actuator strokes. The calculation will therefore become very slow as the number of actuators on the deformable mirror increases. However, in most cases the influence function of one actuator does not spread over the entire pupil and the number of dark-hole points is much smaller than the size of the total image. One can use this property to decrease the number of computations required. In the ideal case where each actuator influence functions do not overlap, a simple solution can be found.

Assume that the influence functions $\sigma_k(u,v)$ do not overlap and the cross product is always zero:

$$\int du\,dv\, \sigma_k(u,v)\sigma_l(u,v) = \delta_{kl}\sigma_k(u,v), \tag{12}$$

where $\delta_{kl}$ is the Kronecker delta function. The effect of each actuator can then be separated. Appendix A.2. shows that the number of calculations can be decreased from $M+1$ Fast Fourier transforms (FFTs) for each iteration to $M+1$ discrete Fourier transforms (DFTs) which take into account only for the pixels affected by the $k^{th}$ actuator. Furthermore the actuator influence functions are square-shaped and not overlapping, then the minimization metric $\chi^2$ is only a function of $f_k$, where $f_k$ is the wavefront amplitude in the focal plane when the $k^{th}$ actuator is set to 1 and all the rest are set to 0. In this case, $\chi^2$ and the least-square gradient are computed by using the following formulae for $A(x_p, y_p)$ the wavefront amplitude in the dark points,

$$A(x,y) = \sum_{k=1}^{M} e^{ia_k} f_k(x,y) - f_0(x,y) \qquad (13)$$

$$\frac{\partial A}{\partial a_k}(x,y) = i e^{ia_k} f_k(x,y). \qquad (14)$$

Note that since $f_k$ is independent of the actuator stroke, these functions can be pre-computed at the beginning of the minimization process. The operations used during the iterative process require no additional Fourier transforms. Therefore we have computed $M + 1$ DFTs for the entire minimization process instead of $M + 1$ FFTs per iteration. Appendix A.3. shows that by assuming square-shaped non-overlapping actuators and with the parameters used, computations of $\chi^2$ and its gradient require about one minute of CPU time on a SPARC2 workstation instead of about 20 h.

## 5. SIMULATION RESULTS

Simulations were performed using square-shaped independent actuator on the *HST* phase map. The simulations were divided into 3 steps. The first one is the initialization of the $\chi^2$ algorithm and the computation of the functions $f_k$, and $f_0$. The second step is to use the pre-computed functions in the Levenberg-Marquardt non-linear minimization process. This results in the optimum actuator strokes needed to minimize the intensity in the dark holes. The third and final step is the application of the actuator strokes and the computation of the image intensity.

A $1024 \times 1024$ array of points was used to sample both the *HST* pupil and the image plane. The *HST* pupil was only 204 pixels across. An image of the phase errors with the spherical aberration removed is shown on Fig. 2. The data was taken from a report by Roddier and Roddier (1990) where phase-retrieval techniques were used on 8-9 WF-PC-1 images during the *HARP* (*Hubble Aberration Retrieval Program*) program to assess the spherical aberration on *HST*. All the following simulations are computed at 0.8 $\mu$m wavelength.

The maximum-strehl algorithm often produces very good results in reducing scatter as explained in Sec. 3.2.. Therefore we used these actuator strokes as a first guess in our $\chi^2$ simulations. The results of the maximum strehl algorithm are displayed on the figures in next section along with the results from dark-hole algorithm.

We tested a number different shapes for the dark holes: a disk, annulus, rectangle, and two symmetric rectangles. For our simulations, we used a deformable mirror with $37 \times 37$ square actuators each $7 \times 7$ pixels in sizes. One of the parameter which defined the shape of dark holes is the sparsity factor. The sparsity factor is the density of points in the dark holes. A sparsity factor of 4 means that only one in every 4 pixels along each axis is used. Table 1 describes the shape used and their parameters. All units are given in pixels. The scale is 14 mas per pixel.

Table 1: Dark-hole parameters used in generating Figures 5–8.

| Figure | Shape | Size | Sparse factor | # dark points |
|--------|-------|------|---------------|---------------|
| Fig. 5 | Disk | radius = 68 px | 5 | 577 |
| Fig. 6 | Rectangle | length = 120 px | 4 | 496 |
|        |           | width = 60 px |   |     |
| Fig. 7 | Symmetrical | length = 60 px | 4 | 544 |
|        | (2 rectangles) | width = 60 px |   |     |
| Fig. 8 | Annulus | inner radius = 8 px | 5 | 568 |
|        |         | outer radius = 68 px |   |    |

Table 2: Summary of simulation results.

| Figure | Shape | Maximum stroke | RMS Stroke | $\chi^2$ |
|--------|-------|----------------|------------|----------|
| Fig. 5 | Disk | 57 nm | 14.4 nm | 40 |
| Fig. 6 | Rectangle | 62 nm | 14.2 nm | 57 |
| Fig. 7 | Symmetrical | 58 nm | 15.9 nm | 23 |
| Fig. 8 | Annulus | 59 nm | 15.6 nm | 16 |

The results are displayed in Fig. 5, 6, 7 and 8. Table 2 summaries the principal features of the dark-hole solution. Corresponding maximum and RMS actuator stroke are listed. The metric ($\chi^2$) is given, showing that the best solution is achieved with the annular dark-hole. In all cases, a level between $10^{-8}$ and $10^{-7}$ times below the intensity of the central star can be reached. The limitations of the dark-hole concept are discussed in the next section. The scattering level increases outside the dark-hole region because the scattered light has been removed from the dark hole region.

To this point we have demonstrated the dark-hole concept using a shearing interferometer for cancelation of scattered light. In reality, the implementation of a shearing interferometer for space application is difficult. The more common method of stray light suppression is the use of a coronograph. We have simulated a coronograph design presented in Malbet, Shao and Yu (1994) and used it in conjunction with a deformable mirror with actuator stroke results computed with annular dark-hole simulations. We modified the entrance pupil of the coronograph to introduce the effect of a deformable mirror correction before the coronograph optics. Figure 9 shows the intensity in the focal plane from a coronograph, with the scattering correction from the deformable mirror. In this case, the use of coronograph does not degrade the performance of the dark-hole algorithm, because a coronograph changes the amplitudes, but not the phases.

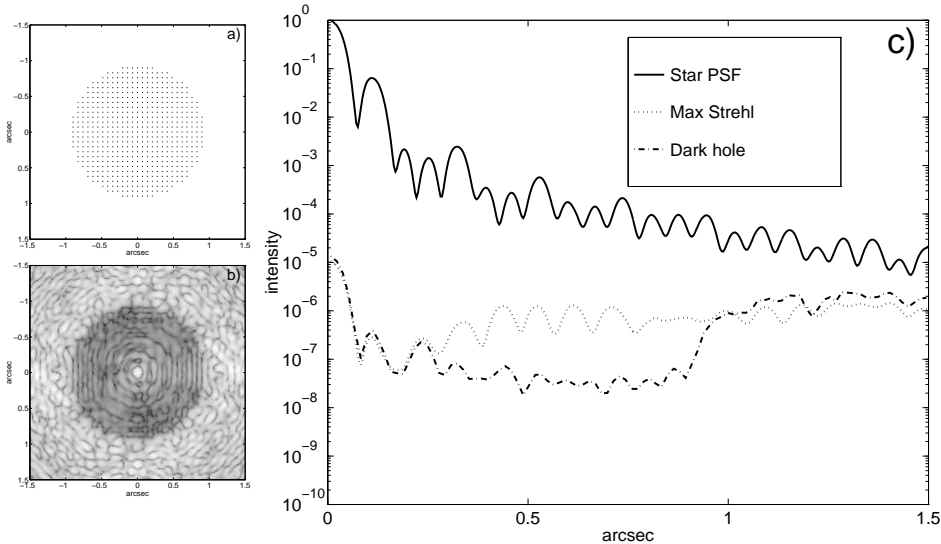

Fig. 5.— Result of the non-linear least-square dark-hole algorithm for a disk-shaped dark-hole pattern. The parameters are: $\lambda = 0.8$ $\mu$m, $1024 \times 1024$ pixel grid, 644 actuators ($37 \times 37$ square grid), 577 dark holes. The maximum actuator stroke is 57 nm and the RMS actuator stroke is 14.4 nm. Panel a) shows the location of the dark-hole points, Panel b) shows the image of the scattering after 20 iterations and Panel c) displays the radial profile of the scattering intensity image averaged azimuthally together with the reference star PSF and the Max Strehl solution.

## 6. LIMITATIONS OF THE DARK-HOLE ALGORITHM

### 6.1. Nyquist Limit

Because of the Nyquist criteria, the amount of scattered light that can be canceled will be limited by the number of actuators across the pupil. With 37 actuators across the pupil, we can correct up to a radius of $18\lambda/D$ from the center of the image. Taking into account the secondary mirror obscuration (0.8m/2.2m), the correction is further limited to a radius of $12\lambda/D$. At $\lambda = 0.8$ $\mu$m that means that we cannot suppress scattered light beyond $0\rlap{.}''9$ from the center of the image.

### 6.2. Dark-Hole Geometry

If the actuator strokes $a_k$ are small ($< \lambda/10$ RMS), $\exp(i\phi)$ can be approximated by $1 + i\phi$. This means that the phase correction will produce an intensity pattern which is symmetric about the origin. Consequently, the use of dark holes which are axisymmetric will result in actuator strokes which are small and easier to realize.

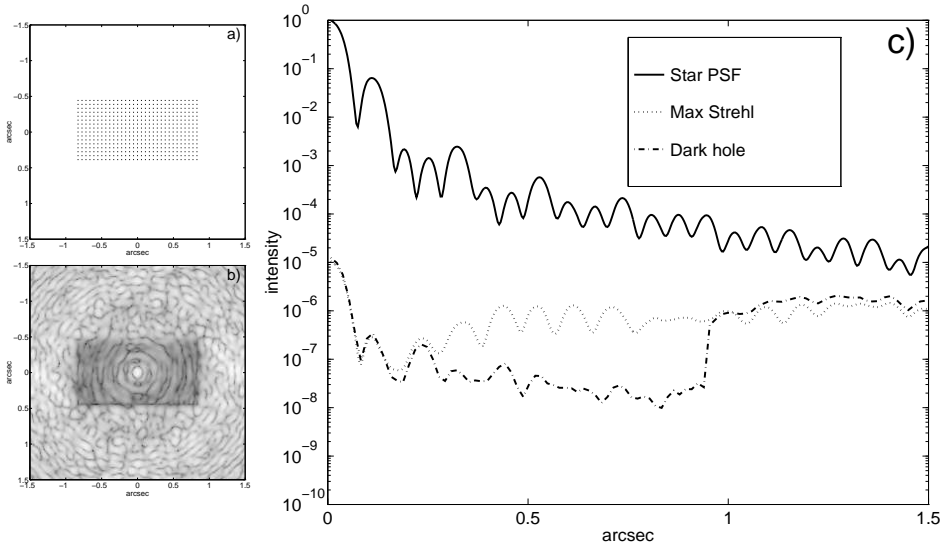

Fig. 6.— Result of the non-linear least-square dark-hole algorithm for a rectangular-shaped dark-hole pattern. The parameters are: $\lambda = 0.8$ $\mu$m, $1024 \times 1024$ pixel grid, 644 actuators ($37 \times 37$ square grid), 496 dark holes. The maximum actuator stroke is 62 nm and the RMS actuator stroke is 14.2 nm. Panel a) shows the location of the dark-hole points, Panel b) shows the image of the scattering after 20 iterations and Panel c) displays the radial profile of the scattering intensity image averaged along the short side of the rectangle together with the reference star PSF and the Max Strehl solution.

There is always some remaining scattering at the center of the image even in dark holes which includes the center (disk or rectangular dark hole). This effect comes from terms of second order ($\propto \phi^2$). When $\phi \ll 1$, then $\exp(i\phi) - 1 = i\phi + \phi^2/2 + ....$ The first order imaginary term can be removed by phase control, but not the second order real term. The only way to suppress it is by amplitude control in the pupil plane. Therefore the deformable mirror cannot suppress this intensity within one PSF width of the center.

We are also constrained by the number of degrees of freedom available to produce dark-hole points. If we had an unobscured pupil, then the entire Nyquist-limited zone could be selected as a dark-hole region. However because of the obscuration from the secondary mirror, there are fewer usable actuators and we cannot decrease the scattering level over the entire Nyquist region.

### 6.3. Wavefront Sensing and Actuator Stroke Accuracy

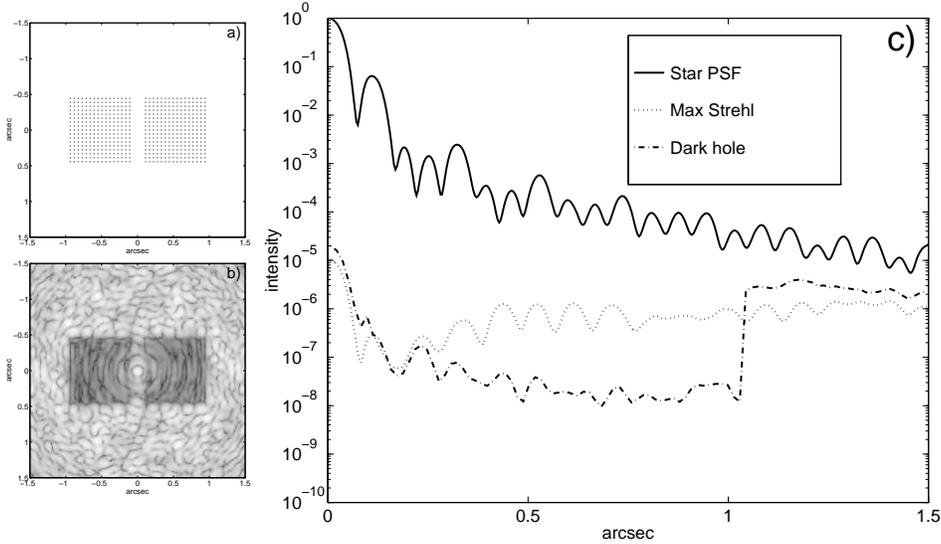

Fig. 7.— Result of the non-linear least-square dark-hole algorithm for a symmetrical-rectangle-shaped dark-hole pattern. The parameters are: $\lambda = 0.8$ $\mu$m, $1024 \times 1024$ pixel grid, 544 actuators ($37 \times 37$ square grid), 577 dark holes. The maximum actuator stroke is 58 nm and the RMS actuator stroke is 15.9 nm. Panel a) shows the location of the dark-hole points, Panel b) shows the image of the scattering after 20 iterations and Panel c) displays the radial profile of the scattering intensity image averaged along the short side of the rectangle together with the reference star PSF and the Max Strehl solution.

In order to perform the dark-hole algorithm, we need an accurate knowledge of the wavefront and an accurate positioning of the actuators. It is important to measure the phase error of the original optical system in order to determine the strokes to apply. We have to know the wavefront with a precision equal to the accuracy we wish to control the wavefront. An extrapolation of *HST* data results in a scattering level of $10^{-8}$ if we can measure the phase errors with an accuracy of $\lambda/2000$ RMS. One way to measure the wavefront is to use a Michelson interferometer, or phase retrieval techniques (Roddier and Roddier 1990). A simpler method would be to use the imaging camera as the wavefront sensor.

To determine the amount of stroke error that can be tolerated, we generated a normal random error on the actuator strokes. Figure 10 summarizes the results obtained with different level of accuracy (0.2 nm, 0.4 nm and 0.8 nm) for the annular dark hole. Table 3 gives a summary of the scatter suppression for the zone located between 0.3″ and 0.7″. Errors less than 0.2 nm do not significantly change the scattering level of the original result ($3 \times 10^{-8}$). This result constraints the specification for the actuator stroke accuracy of the deformable mirror. In our case, the stroke accuracy needs to be better than 0.4 nm to achieve a scattered light level of the order of $3 \times 10^{-8}$.

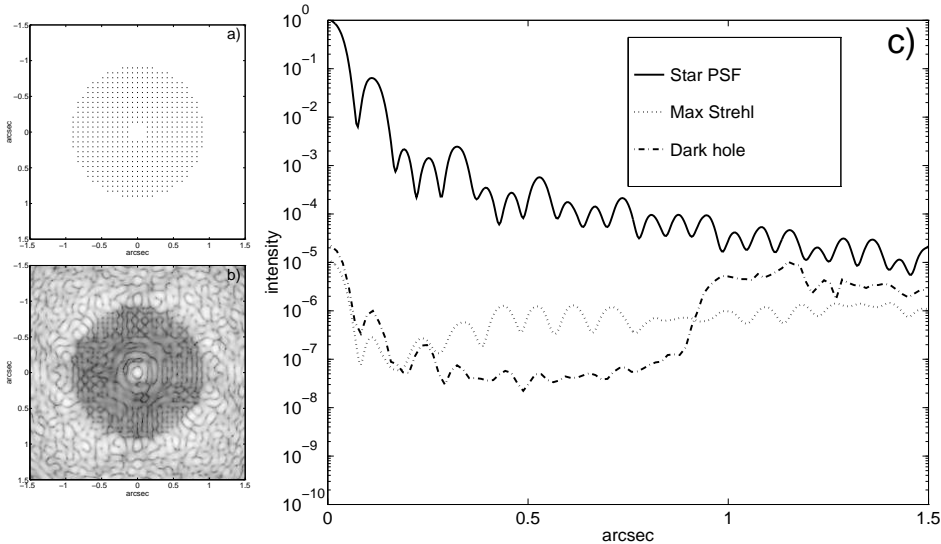

Fig. 8.— Result of the non-linear least-square dark-hole algorithm for a annular-shaped dark-hole pattern. The parameters are: $\lambda = 0.8$ $\mu$m, $1024 \times 1024$ pixel grid, 644 actuators ($37 \times 37$ square grid), 568 dark holes. The maximum actuator stroke is 59 nm and the RMS actuator stroke is 15.6 nm. Panel a) shows the location of the dark-hole points, Panel b) shows the image of the scattering after 20 iterations and Panel c) displays the radial profile of the scattering intensity image averaged azimuthally together with the reference star PSF and the Max Strehl solution.

We would like to point out that the quality of the deformable mirror is not a very important issue, because the mirror figure errors with spatial period smaller than the actuator size will produce scattered light beyond the Nyquist limit. However the higher the surface quality of the deformable mirror, the easier the suppression of the scattered light.

### 6.4. Wavelength Sensitivity

The dark-hole algorithm computes the optimum actuator spacing at a particular wavelength. For a fixed actuator configuration, when the wavelength changes two things occur. First the depth of the dark hole changes proportional to $\lambda^2$, longer wavelength producing greater suppression since the effective phase error is decreased. Secondly, the distance of the dark hole from the center of the image plane changes linearly with the wavelength. In order for the dark-hole algorithm to function over a broad wavelength band, it is essential that there is significant overlap in the dark-hole regions over the entire observation spectral band. Consequently it is important to cancel stray light in regions as close to the center of the field of view as possible.

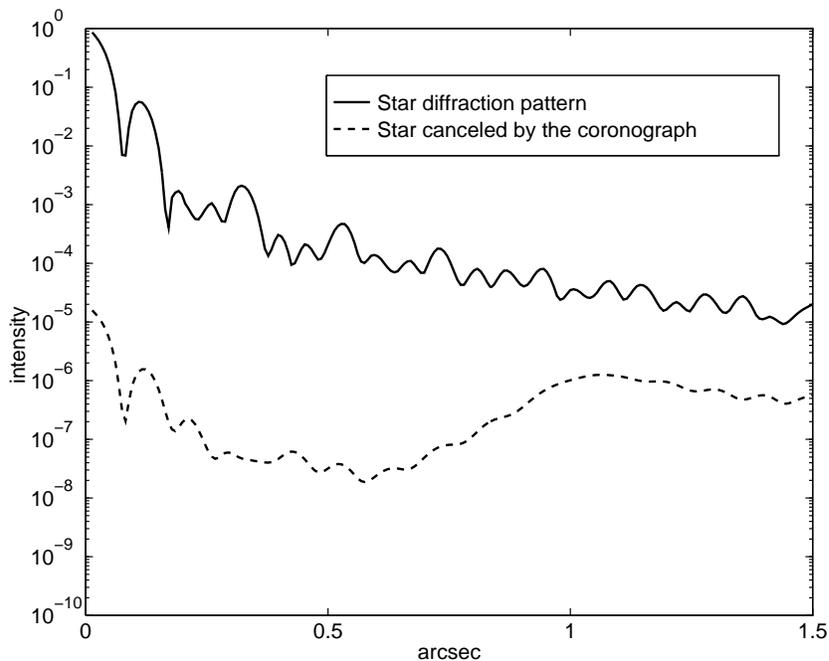

Fig. 9.— Scattering level with a deformable mirror and a coronograph (see text for details)

Figure 11 shows results for the case of three different wavelengths. The optimization has been performed at 0.8 $\mu$m and the resulting intensity is displayed at 0.6 $\mu$m, 0.8 $\mu$m and 1 $\mu$m.

## 7. SPACE DEFORMABLE MIRROR

The space based coronograph will require a high density deformable mirror with low stroke to correct for wavefront errors at low spatial scales. Because of packaging constraints, the image of the pupil on the deformable mirror will not be very large. As an example, a design for the *HST* advanced camera (Macenka, private communication) resulted in a 2.5 cm diameter for the re-imaged pupil. With actuator spacings of 1 mm, this would result in 490 actuators across the mirror. The deformable mirror would then be capable of only correcting errors of the *HST* primary at spatial scales larger than 9.6cm.

### 7.1. Space Active Optics vs. Ground-Based Adaptive Optics

While the adaptive optics (AO) concept is the same for ground and for space, its implementation in space and on the ground are very different. Ground-based AO systems are designed to remove the time-varying phase errors introduced by atmospheric turbulence. Space AO has an entirely different goal, namely, to correct the fabrication errors in the large light-collecting optics and to thereby reduce the scattered light caused by those imperfections. Table 4 shows the relevant parameters of space vs. ground AO systems.

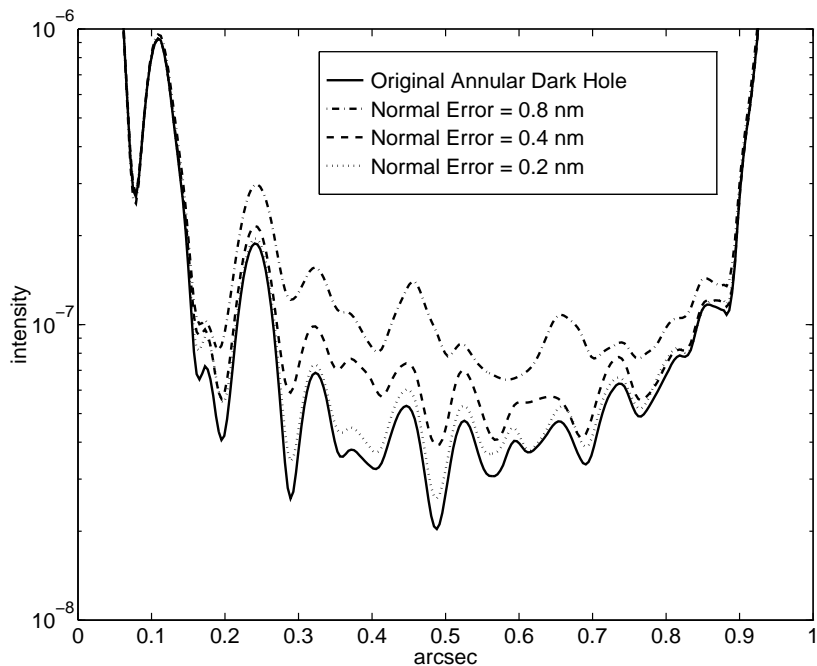

Fig. 10.— Scattering level for an annular dark hole with different actuator stroke accuracies. The simulation for a perfect actuator (Fig. 8) is represented by the solid line. Errors of different magnitude (0.8 nm, 0.4 nm, 0.2 nm) are represented by the broken lines. An error less than 0.2 nm does not significantly change the solid-line result.

Instead of correcting the wavefront for turbulence 100 times a second, space AO is a one-time-only (or e.g. low bandwidth) procedure. In space there is no need for a separate wavefront sensor since the science camera can be used to detect the wavefront error. Phase retrieval techniques are then used on the science data to determine mirror correction. Perhaps the biggest difference is in the active mirror. On the ground the active mirror must remove the effects of atmospheric turbulence which can require up to 10 $\mu$m of optical path for 8-m and 10-m telescopes. In space because the primary and secondary mirror is well polished, the required stroke is very small, 50 nm at most. This can be implemented with a new generation of active mirrors for space which can be built at low cost.

Table 3: Levels obtained with different actuator stroke precisions.

| Mean Stroke Error | Mean Scattering Level |
|---|---|
| 1 nm | $1.1 \times 10^{-7}$ |
| 0.8 nm | $8 \times 10^{-8}$ |
| 0.6 nm | $7 \times 10^{-8}$ |
| 0.4 nm | $6 \times 10^{-8}$ |
| 0.2 nm | $4 \times 10^{-8}$ |
| 0.1 nm | $3 \times 10^{-8}$ |

Table 4: Space active optics vs. ground adaptive optics specifications

| Parameter | Ground (D=10m, $r_0$=20cm) | Space |
|---|---|---|
| # actuators | 2500 | 300-1200 |
| actuator stroke | 10 $\mu$m | 50 nm |
| actuator resolution | 10 nm | 0.1 nm |
| servo bandwidth | 50-200 Hz | <0.001 Hz |
| wavefront sensor | Hartman/curvature etc. | science camera (using phase retrieval) |

### 7.2. Integrated/Monolithic Active Mirrors

Active mirrors for ground-based telescopes come in several flavors. Some are based on stacks of PMN/PZT's with face sheets. Others are based on PZT bimorphs. Yet others are based on PZT/PMN tubes. The earliest deformable mirror was based on a monolithic piece of PZT with an electrode pattern deposited on the PZT to define the actuators. This type of active mirror is no longer used because the stroke of a single PZT layer is insufficient to correct the $\approx 10$ $\mu$m effects of atmospheric turbulence. For space applications, however, this single layer of PMN/PZT is sufficient to change the optical path by $\approx 50$ nm. Figure 12 shows the concept for a monolithic active mirror, where photolithographic techniques are used to define the size and number of actuators. Another simplification of space AO is in the electronics that drive the active mirror. High bandwidth active mirrors require one high voltage amplifier per actuator. PMN/PZT actuators, however, are totally capacitive and require power only when the actuator is moved. No power is required if the mirror is asked to "hold" a particular position. Hence, a single amplifier with a multiplexer can be used to drive the whole active mirror in space.

### 7.3. Deformable Mirror for Space

Because of the high actuator densities required, the deformable mirror will require an integrated method of fabrication. In the design concept shown in Fig. 12 (right panel), a single piezoelectric wafer sandwiched between electrodes is used to provide the force necessary to deform the mirror. Because of the initial quality of the *HST* optics the active mirror has a low stroke requirement ($\approx \pm 50$ nm). Consequently, the PZT wafer does not need to be diced. An electrode pattern is deposited on the bottom layer of the PZT wafer. This electrode pattern is used to drive local regions in the PZT wafer. Another metal layer is deposited on the top of the PZT and serves as the ground plane. The electrode pattern will be fabricated using photolithographic techniques and may require multiple layers to run signal lines to the periphery of the device. In addition, the fabrication procedures must be performed at temperatures which do not depole the PZT wafer and must result in a flat device so print-through effects do not show up on the top surface of the integrated PZT wafer. The whole sandwich assembly is bonded on a glass substrate for structural support. A thin glass surface is then glued on the top of the PZT sandwich. In the final step, this surface is super-polished and coated to provide an extremely smooth reflecting surface for the mirror.

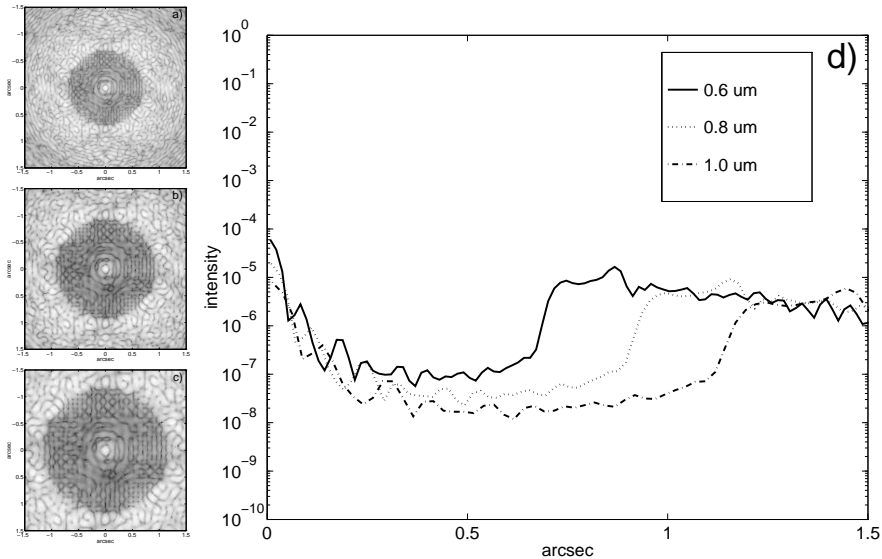

Fig. 11.— Scattering level for the annular dark hole solution of Fig. 8 at three different wavelengths 0.6 $\mu$m (a), 0.8 $\mu$m (b), 1.0 $\mu$m (c). The actuator strokes were optimized for $\lambda = 0.8$ $\mu$m. Then the scattering intensity image were computed for different wavelengths. Panel (d) displays the radial profile of the scattering intensity images averaged azimuthally at the 3 different wavelengths.

## 8. CONCLUSION

In this paper, we have shown that scattered light is a limiting factor for high dynamic range imaging. The scattering can be suppressed in specific regions called dark holes by using a deformable mirror in a pupil plane. The scattering is due to small error figures in the mirrors of telescopes. Adaptive optics can be used to cancel the scattered light in the focal plane by using a dark-hole algorithm which drives the actuators of the deformable mirror in order to suppress the light at low spatial frequency and lower the scattering level near the optical axis (center of the image). Simulations performed with a range of different dark hole shapes (annular, rectangular, symmetric, disk) have resulted in scattering level as low as $2 \times 10^{-8}$ times the peak intensity of the bright star. We have also presented a concept for a space-based deformable mirror for scattered light suppression. We are currently working on a laboratory experiment to demonstrate the dark-hole concept.

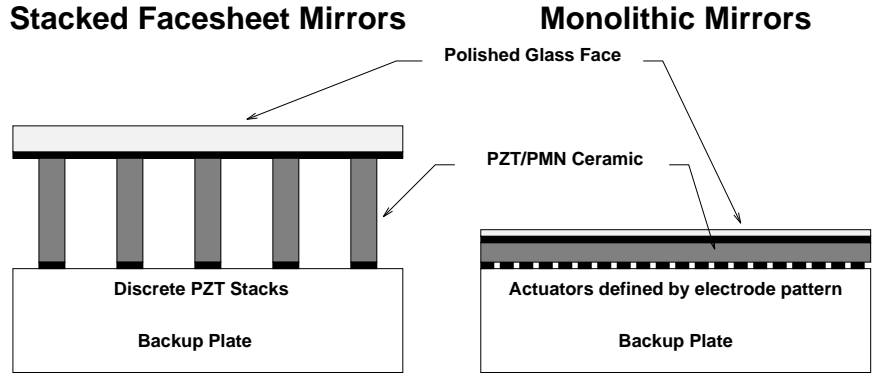

Fig. 12.— Comparison between a ground-based stacked facesheet mirror (left) and a space-based monolithic mirror (right). The stacked facesheet mirror has larger strokes (larger than 10 $\mu$m), needs space between actuators for large motions and the quality of the facesheet figure depends on the thickness of the facesheet. The monolithic mirror has smaller strokes (about 0.1 $\mu$m), has no gaps between the actuators, is made with one piezo layer (no stacks) and the solid mirror can be polished like ordinary optics.

We would like to thank Claude Roddier for providing the *Hubble Space Telescope* wavefront data. This work was performed at the Jet Propulsion Laboratory, California Institute of Technology, under a contract with the National Aeronautics and Space Administration (NAS7-1260 and SPI 1-02-4A)

(15)

(1)

## A. DARK-HOLE NON-LINEAR LEAST-SQUARES ALGORITHM

The dark-hole algorithm uses the following metric

$$\chi^2 = \sum_{p=1}^{P} |A(x_p, y_p)|^2, \tag{A2}$$

where $A(x,y)$ is the wavefront amplitude in the focal plane and $P$ is the number of dark-hole points. In order to minimize $\chi^2$, we use a Levenberg-Marquardt non-linear least-square routine. This method converges toward a solution by calculating $\chi^2$ and its gradient with respect to the unknown parameters (the actuator strokes $a_k$). A normal implementation of the method would require $M+1$ Fourier transforms for each iteration. In this appendix, we show that by making a number of simplifying assumptions we can reduce the number of operations required to compute the actuator strokes.

Section A.1. compares the number of operations needed in a Fast Fourier Transform (FFT) implementation compared to those needed in a Discrete Fourier Transform (DFT) implementation. Section A.2. shows that by assuming non-overlapping influence functions, the computation of $\partial A/\partial a_k$ depends only on the $k^{th}$ actuator. Finally in Sect. A.3. we show that in the case of square-shaped non-overlapping influence functions the minimization process does not require Fourier transforms calculation at each iterative step, but only at the initial step of the non-linear least-square algorithm.

### A.1. Discrete vs. Fast Fourier Transform

For our particular algorithm, we only need to compute the intensity at $P$ dark-hole points. In typical situations $P$ is much less than the total number of points in the image. In this case, the number of operations needed to implement a DFT can be less than performing a full FFT. As an example, our simulations use $N = 1024^2$ array and a configuration with $M = 644$ actuators each having $n = 7^2$ pixels, and $P = 568$ dark holes (in the case of the annular dark hole). A FFT requires $N \times \log_2 N = 2 \times 10^7$ operations. For a DFT, we need to know the number of input and output points. Since we padded the pupil with zeroes by a factor 4, the number of input points is only $M \times n$ (about 33 times smaller than $N$). The number of output points is the number of dark holes, $P$. Therefore, computing the DFT for the $P$ dark-hole points requires $M \times P \times n = 1.7 \times 10^7$ operations.

The gain is not outstanding, but next section shows that the number of input points decreases if the actuator influence functions are independent. In this case, computing DFTs becomes then much faster than computing FFTs.

### A.2. Independent Actuator Influence Functions

Equation (5) represents the stray light intensity on the detector and can be written as $I(x,y) = |A(x,y)|^2$ where

$$A(x,y) = \text{FT} \left\{ \Pi(u,v) \left( e^{i\phi_0(u,v)} e^{i \sum_{k=1}^{M} a_k \sigma_k(u,v)} - 1 \right) \right\}. \tag{A3}$$

is the wavefront amplitude in the image plane. $\phi_0(u,v)$ is the actual phase map of the pupil, and $a_k$ and $\sigma_k(u,v)$ are the stroke and influence function of the $k^{th}$ actuator. $\Pi(u,v)$ is the pupil transmission function.

If we assume that the actuator influence functions are independent, i.e do not overlap, then we get the relation described in Eq. (12). Note that

$$\exp\left( i \sum_{k=1}^{M} a_k \sigma_k(u,v) \right) = 1 + \sum_{n=1}^{\infty} \frac{\left( i \sum_{k=1}^{M} a_k \sigma_k(u,v) \right)^n}{n!} \tag{A4}$$

$$= 1 + \sum_{n=1}^{\infty} \frac{\sum_{k=1}^{M} (i a_k)^n \sigma_k(u,v)^n}{n!} \tag{A5}$$

$$= 1 + \sum_{k=1}^{M} \left( e^{i a_k \sigma_k(u,v)} - 1 \right). \tag{A6}$$

Hence

$$\frac{\partial A}{\partial a_k}(x,y) = i \text{FT} \left\{ \Pi(u,v) \sigma_k(u,v)) e^{i(\phi_0 + a_k \sigma_k(u,v))} \right\}. \tag{A7}$$

Then the derivative of $A(x,y)$ with respect to the $k^{th}$ actuator depends only on the stroke and influence function of that actuator. If the actuator influence function has a limited size compared to the pupil the number of operations required by a DFT is much smaller. In our example, the influence functions are not overlapping and there are only $P \times n = 2.7 \times 10^4$ operations required. Thus a large gain over performing an FFT for each $\partial A / \partial a_k$.

### A.3. Square-Shaped Actuator Influence Functions

Consider now the case of square-shaped actuator influence functions

$$\sigma_k(u,v) = \begin{cases} 1 \text{ if } (u,v) \text{ is on actuator } k, \\ 0 \text{ if } (u,v) \text{ is not on actuator } k. \end{cases} \tag{A8}$$

This kind of actuator function does not overlap and is independent. Therefore Eq. (A3) can be rewritten

$$A(x,y) = \text{FT} \left\{ \Pi(u,v) \left( e^{i\phi_0(u,v)} \left[ 1 + \sum_{k=1}^{M} \left( e^{i a_k} - 1 \right) \sigma_k(u,v) \right] - 1 \right) \right\}, \tag{A9}$$

and

$$\frac{\partial A}{\partial a_k}(x,y) = ie^{ia_k}\text{FT}\left\{\sigma_k(u,v)\Pi(u,v)e^{i\phi_0(u,v)}\right\}. \tag{A10}$$

Furthermore if we assume

$$\Pi(u,v)\sum_{k=1}^{M}\sigma_k(u,v) = \Pi(u,v), \tag{A11}$$

i.e. if the actuators are next to each other with no gaps and that they populate the entire pupil, then

$$A(x,y) = \sum_{k=1}^{M} e^{ia_k}\text{FT}\left\{\sigma_k(u,v)\Pi(u,v)e^{i\phi_0(u,v)}\right\} - \text{FT}\{\Pi(u,v)\} \tag{A12}$$

For the minimization process, we define the following functions

$$f_0(x,y) = \text{FT}\{\Pi(u,v)\} \tag{A13}$$

$$f_k(x,y) = \text{FT}\left\{\sigma_k(u,v)\Pi(u,v)e^{i\phi_0(u,v)}\right\} \tag{A14}$$

which are the ideal point spread function and the impulse functions of each actuator respectively. Then for the $\chi^2$ minimization the wavefront amplitude in the focal plane and its partial derivatives are :

$$A(x,y) = \sum_{k=1}^{M} e^{ia_k} f_k(x,y) - f_0(x,y) \tag{A15}$$

$$\frac{\partial A}{\partial a_k}(x,y) = ie^{ia_k} f_k(x,y). \tag{A16}$$

Since both $f_k$ and $f_0$ are independent of $a_k$, we only need to compute $M+1$ reduced DFTs at the beginning of the minimization process. Thus the number of DFTs is independent of the number of iterations required to converge toward a solution, thereby increasing the speed of the least-square minimization and making the computation for a large number of actuators possible.

In our simulations, typically $N_i = 20$ iterations were necessary. A straight forward method using FFTs at each iteration requires $N_i \times (M+1) \times (N\log_2 N)$ operations, i.e. in our case, $3 \times 10^{11}$ operations to compute $A(x,y)$ and its derivatives. With the previous assumptions and by using DFTs, we need only $(M+1) \times P \times n$ operations, i.e. $1.8 \times 10^7$. There is gain by a factor 15,000 in computation time. On our SPARC-2 workstation, it takes about 45 s of CPU time to calculate the full DFT implementation for the entire process. With the FFT implementation, it would have taken over 200 h to calculate the $\chi^2$ and its derivatives.

---